\documentclass[journal,twoside,web]{ieeecolor}
\usepackage{generic}
\usepackage{cite}
\usepackage{amsmath,amssymb,amsfonts}
\usepackage{algorithmic}
\usepackage{graphicx}
\usepackage{textcomp}
\newtheorem{Theorem}{Theorem}
\newtheorem{Definition}{Definition}
\newtheorem{Remark}{Remark}
\newtheorem{Lemma}{Lemma}

\newtheorem{Proposition}{Proposition}
\newtheorem{Assumption}{Assumption}

\def\BibTeX{{\rm B\kern-.05em{\sc i\kern-.025em b}\kern-.08em
    T\kern-.1667em\lower.7ex\hbox{E}\kern-.125emX}}
\begin{document}
\title{Understanding the Capability of PD Control for Uncertain Stochastic Systems}
\author{Cheng~Zhao and Yanbin~ Zhang
\thanks{This paper was supported by the National Natural Science Foundation of China under Grant No. 12288201. (Corresponding author: Cheng Zhao.)\\
\indent Cheng Zhao is with the Key Laboratory of Systems and Control,
Academy of Mathematics and Systems Science, Chinese Academy of
Sciences, Beijing 100190, China(e-mail: zhaocheng@amss.ac.cn).\\
\indent Yanbin Zhang is with Beijing SciMath International Education Technology Co., LTD, China(e-mail: x15yanbin.zhang@gmail.com). }}
%This paragraph of the first footnote will contain the date on
%which you submitted your paper for review. It will also contain support
%information, including sponsor and financial support acknowledgment. For
%example, ``This work was supported in part by the U.S. Department of
%Commerce under Grant BS123456.'' }
%\thanks{The next few paragraphs should contain
%the authors' current affiliations, including current address and e-mail. For
%example, F. A. Author is with the National Institute of Standards and
%Technology, Boulder, CO 80305 USA (e-mail: author@boulder.nist.gov). }
%\thanks{S. B. Author, Jr., was with Rice University, Houston, TX 77005 USA. He is
%now with the Department of Physics, Colorado State University, Fort Collins,
%CO 80523 USA (e-mail: author@lamar.colostate.edu).}
%\thanks{T. C. Author is with
%the Electrical Engineering Department, University of Colorado, Boulder, CO
%80309 USA, on leave from the National Research Institute for Metals,
%Tsukuba, Japan (e-mail: author@nrim.go.jp).}}

\maketitle

\begin{abstract}
%Undoubtedly, most of the practical control systems are nonlinear with various uncertainties and random disturbances.
In this article, we focus on the global stabilizability problem for a class of second order uncertain stochastic control systems, where both the drift term and the diffusion term are nonlinear functions of the state variables and the control variables. We will show that the widely applied proportional-derivative(PD) control in engineering practice has the ability to globally stabilize such systems in the mean square sense, provided that the upper bounds of the partial derivatives of the nonlinear functions satisfy a certain algebraic inequality. It will also be proved that the stabilizing PD parameters can only be selected from a two dimensional bounded convex set, which is a significant difference from the existing literature on PD controlled uncertain stochastic systems. Moreover, a particular polynomial on these bounds is introduced, which can be used to determine under what conditions the system is not stabilizable by the PD control, and thus demonstrating the fundamental limitations of PD control.
\end{abstract}


\section{Introduction}
Feedback is a basic concept in automatic control, which has had a revolutionary influence in practically all areas. Its primary objective is to reduce the effects of the plant uncertainty on the desired control performance(e.g., stability, optimality of tracking, etc).
%In fact, Dealing with uncertainty is a fundamental issue in control theory.
Plenty of control methods have been developed for dealing with uncertainties over the past sixty years, such as  adaptive control \cite{krstic}, robust control\cite{zhou1998}, active disturbance rejection control \cite{han2009,chen2020} and sliding mode control, etc. However, the classical proportional-integral-derivative(PID) control, perhaps the most basic form of feedback, has been at the heart of control engineering practice for several decades\cite{Astrom2000}. In fact, the PID control is used in more than 90\% of industrial processes \cite{Astrom1995}. One may naturally believe that such basic controller has been deeply understood in both theory and practice. However, as mentioned  in \cite{o2006pi}, many practical PID loops are poorly tuned, and there is strong evidence that its rationale remains to be unclear.

Recently, the PID control has attracted more and more attention from the research community. For example, the stabilization problems of PD(or PID) controlled linear systems with time-delay are investigated, see e.g., \cite{li2020,ma2019,Silva,chen2022}. There are also abundant works on PD controlled mechanical systems(see e.g., \cite{Borja,kelly,Takegaki}), among which \cite{Takegaki} is probably the most notable, where a PD controller was constructed to globally stabilize fully actuated robot manipulators. For more general class of nonlinear uncertain systems without special structures, some rigorous mathematical investigations have been made on the theory and design of PID in recent years(see, e.g.,\cite{zhao2017,Zhang2019,zhao2020,zhao2021}). For instance, it has been shown that for a class of second order single-input-single-output(SISO) affine nonlinear system, one can select the three PID parameters to globally stabilize the closed-loop system and at the same time to make the output of the controlled system converge to any given setpoint, provided that the partial derivatives of the system nonlinear functions are bounded \cite{zhao2017}. Extensions to MIMO non-affine systems without stochastic disturbances are discussed in \cite{zhao2020}.
%It is worth mentioning that most of the theoretical investigations on the analysis and design of PID control are focused on deterministic system, while stochastic systems are becoming extensively used as realistic models of physical phenomena since random disturbances and measurement noises are inevitable in most practical systems.
%In fact, the importance of taking disturbances into account has been realized by practitioners from the beginning of the development of control theory\cite{Åström}.
More recently, the authors investigate the performance and design of PID control for non-affine stochastic systems in \cite{zhang2021}, where the diffusion term does not depend on the control input.

As a special case of PID, the PD controller has also attracted many scientists and scholars, see \cite{Takegaki,kelly,ma2019,zhao2017,zhang2021}. In order to understand the mechanism of the linear PD control, it is of vital importance to take nonlinearity, uncertainty and randomness into consideration. Moreover, efforts must be taken to investigate the limitations of PD control in a general framework. But, to the best of the our knowledge, these issues have not been fully explored. In this article, we are devoted to this fundamental problem by considering a basic class of MIMO stochastic nonlinear uncertain systems, where both the drift term and the diffusion term are functions of the state variables and the control variables. The main contributions are summarized as follows:\\
1) We have shown that the PD control has the ability to globally stabilize such systems in mean square, if the upper bounds of the partial derivatives of the nonlinear functions satisfy a certain algebraic inequality. Moreover, a particular polynomial is introduced, which can be used to determine under what conditions the system is not stabilizable by PD control, and thus demonstrating the fundamental limitations of PD control.
\\
2) Open and bounded parameter sets for the controller gains are also constructed, which are based on some knowledge of both the drift and diffusion functions. Besides, it will be shown that the PD parameters cannot be chosen arbitrarily large, which is also a significant difference from the existing literature on PD controlled nonlinear uncertain systems, see e.g. \cite{Khalil},\cite{zhao2017},\cite{cong2017},\cite{zhang2021}.
%\\
%3) For a class of linear stochastic systems, a necessary and sufficient condition on the upper bounds of the partial derivatives of the functions is also provided, under which the system is stabilizable by the PD control.
	
The rest of this article is organized as follows.
In  Section II,  we will introduce the mathematical formulation. The main results are presented in  Section III.   Section IV contains the proofs of the main theorems. Section V will conclude the article with some remarks.

\section{Mathematical Formulation}
\subsection{Notations and Definitions}
Let $\mathbb{R}^n$ be the $n$-dimensional Euclidean space, $~\mathbb{R}^{m\times n}$ be the space of $m\times n$ real matrices. Denote $\|x\|$  as the Euclidean norm of a vector $x$, and $x^{\mathsf{T}}$ as the transpose of a vector or matrix $x$.
The  norm of a matrix $P \in \mathbb{R}^{m\times n}$  is defined by $\|P\|=\sup_{x\in\mathbb{R}^n, \|x\|=1} \|Px\|$.  For a square matrix $P\in\mathbb{R}^{n\times n}$,  denote $P^{\text{sym}}:=(P+P^{\mathsf{T}})/2$ as the symmetrization of $P$, and $\text{tr}(P)$ as the trace of $P$. For a symmetric matrix $S\in \mathbb{R}^{n\times n}$, we denote $\lambda_{\text{min}}(S)$ and $\lambda_{\text{max}}(S)$ as the smallest and the largest eigenvalues of $S$, respectively. For two symmetric matrices $S_1$ and $S_2$ in $\mathbb{R}^{n\times n}$, the notation $S_1>S_2$ implies that  $S_1-S_2$ is a positive definite matrix; $S_1\geq S_2$ implies  that $S_1-S_2$ is a positive semi-definite matrix.
 Let $C^{1}(\mathbb{R}^{n},\mathbb{R}^{m})$ be the space of continuously differentiable functions from $\mathbb{R}^{n}$ to $\mathbb{R}^m$, denoted as   $C^{1}(\mathbb{R}^{n})$ for simplicity when $m=1$. Denote $C^{k}(\mathbb{R}^{n})$ as the space of functions from $\mathbb{R}^{n}$ to $\mathbb{R}$ with $k-$times continuous partial derivatives.

%Given a function $V(x,t)\in C^{2}(\mathbb{R}^{n}\times\mathbb{R}^{+})$ associated with the above SDE. The differential operator $L$ acting on $V$ is defined by
%	\[
%	LV(x,t)=\frac{\partial V}{\partial t}(x,t)+\frac{\partial V}{\partial x}f(x)+
%\frac{1}{2}\text{tr}\Big\{\sigma^{\tau}(x)\frac{\partial^{2}V}{\partial x^{2}}\sigma(x)\Big\},
%	\]
%where  $\frac{\partial^2 V}{\partial x^2}=( \frac{\partial^2 V}{\partial x_{i} \partial x_{j}})$ is the Hessian matrix of $V$.

\subsection{The Control System}
Consider a basic class of second order nonlinear uncertain stochastic control system:
\begin{align}\label{sysn1}
\mathrm{d} x_1=&x_2 \mathrm{d} t\nonumber\\
\mathrm{d} x_2=&f(x_1,x_2,u)\mathrm{d} t+g(x_1,x_2,u) \mathrm{d} B_t,
\end{align}
where $x_1,x_2\in\mathbb{R}^n$ are the system state vector, $u\in\mathbb{R}^n$ is the control input, $B_t\in\mathbb{R}^{1}$ is an one-dimensional
standard Brownian motion, and  the nonlinear functions  $f$ and $g$ belong to $C^1(\mathbb{R}^{3n},\mathbb{R}^{n})$, which may contain unknown dynamics.

%We remark that many practical dynamical systems can be described by the model (1) via the Newton's second law in mechanics or some fundamental physical laws in electromagnetics. For example, the spring oscillator, inverted pendulum, damped vibration, etc.  Moreover, the well-known Langevin equation can also be described by the model (1).

%The control objective is to design a feedback controller to globally stabilize the system (\ref{sysn1}) in the mean square in the presence of uncertainty, namely find $u=u(x_1,x_2)$ such that the solution of (\ref{sysn1}) satisfies
%\begin{align*}
%\lim_{t\to\infty} \mathbb{E}\left[\|x_1(t)\|^2+\|x_2(t)\|^2\right]=0,~~~ \forall x_1(0),x_2(0)\in\mathbb{R}^n,
%\end{align*}
%under the condition $f$ and $g$ are uncertain.

In this article, we aim to study the capability together with a design method of the classical PD control(also abbreviated as ``the PD control''):
\begin{align}\label{pdd}u(t)=k_pe(t)+k_d\dot e(t),~~e(t)=y^*-x_1(t),\end{align}
where $y^*\in\mathbb{R}^n$ is the setpoint, $e(t)$ is the regulation error, $k_p$ and $k_d$ are the PD parameters.

The objective is to design suitable PD parameters to globally stabilize and regulate system (1) in mean square, i.e.,
\begin{align}\label{3.0}\lim_{t\to\infty} \mathbb{E} \left[\|e(t)\|^2\!+\!\|\dot e(t)\|^2\right]\!=0, ~\forall (x_1(0),x_2(0))\in\mathbb{R}^{2n},
\end{align}
where $\mathbb{E}$ denotes the expectation of a random variable.

We first introduce a basic assumption that will be used throughout the article.
\begin{Assumption} The setpoint $y^*\in\mathbb{R}^n$ is an equilibrium of the uncontrolled stochastic system (1). To be precise,
\begin{align}\label{30} f(y^*,0,0)=0,~~~g(y^*,0,0)=0.\end{align}

\end{Assumption}
\vskip 0.25cm
%From Assumption 1, we know that both the drift and the diffusion function are required to be vanished
%at the setpoint. In fact,
\noindent It is worth noting that Assumption 1 is necessary for the existence of $(k_p,k_d)$ to achieve the control objective (3).   Specifically, we have the following proposition.

\begin{Proposition}
Consider the PD controlled system (1)-(2),
where the functions  $f$ and $g$ are  Lipschitz continuous.
Suppose that there exist some PD parameters $k_p$, $k_d$ and some $(x_1(0),x_2(0))\in\mathbb{R}^{2n}$, such that the solution of the closed-loop system satisfies
$\lim_{t\to\infty} \mathbb{E}\big [\|e(t)\|^2+\|\dot e(t)\|^2\big]=0,$
then $f(y^*,0,0)=0$ and $g(y^*,0,0)=0$.
\end{Proposition}
The proof of Proposition 1 is given in  Appendix B.
%The main target of this article is to answer how much uncertainty in $(f,g)$ can be dealt with by the PD control law (2).

Note that both $f(\cdot)$ and $g(\cdot)$ are uncertain functions, we need to find a suitable measure to quantitatively describe the size of uncertainty.  The upper bounds of the partial derivatives of the uncertain functions, which reflect the ``sensitivity" to their variables, are a natural choice for such measurement, see e.g. \cite{xie2000},\cite{zhao2017}. In addition, in order to enable the input signal to affect the state of the controlled system,  the control gain matrix $\frac{\partial f}{\partial{u}}$ should not vanish. These natural intuitions inspired us to introduce the  following assumption.

\begin{Assumption} The drift function $f(\cdot)\in \mathcal{F}_{L_{1},L_{2}}$, where
\begin{align}\mathcal{F}_{L_{1},L_{2}}:=
\Big \{ f: \|\frac{\partial f}{\partial x_{i}}\|\leq L_{i};~\frac{\partial f}{\partial u}\geq I_n,~ \forall x_1,x_2,u\Big\},
\end{align}
where $L_1$, $L_2$ are positive constants, $I_n$ is the $n\times n$ identity matrix, $\frac{\partial f}{\partial x_{i}}$, $\frac{\partial f}{\partial u}$ are the $n\times n$ Jacobian of $f$ with respect to $x_i$ and $u$, respectively. Moreover, the diffusion function $g(\cdot)$ belongs to
\begin{align}\mathcal{G}_{N_{1},N_{2},M}\!:=\!\Big \{ g:
\|\frac{\partial g}{\partial x_{i}}\|\leq N_{i},\|\frac{\partial g}{\partial u}\|\leq \!M,~\! \forall x_1,x_2,u\Big\},
\end{align}
where the constants $N_1$, $N_2$ are positive and $M$ is nonnegative.
\end{Assumption}

%We now give some explanations for the uncertain function spaces.  First, it is worth mentioning that no specific structural information is assumed in this article except for some prior information on the upper bounds of the partial derivatives. Since the classical PID is a linear feedback, the boundedness of the partial derivatives $\frac{\partial f}{\partial x_i}$($i=1,2$)(or the linear growth condition) appears to be necessary in general for global results \cite{zhao2016}.
%Moreover, we remark that the five constants can be used to describe the system uncertainty quantitatively, since the ``size'' of the uncertain function spaces  will increase as long as any of the five constants increases.

Next, we introduce the following definition.
\begin{Definition}
We say that the uncertain stochastic system (1) is (globally) \emph{stabilizable} by the PD control (2), if there exist some PD parameters $(k_p,k_d)\in\mathbb{R}^2,$ such that the control performance (\ref{3.0}) is satisfied for all functions  $f$ and $g$ that satisfy Assumptions 1 and 2. Otherwise, we say that system (1) is not stabilizable by the PD control (2).
\end{Definition}
\begin{Remark} It is known that, if system (1) has the following special form(the diffusion term does not depend on $u$):
\begin{align}\label{nou}
\mathrm{d} x_1=&x_2 \mathrm{d}t\nonumber\\
\mathrm{d}x_2=&f(x_1,x_2,u)\mathrm{d}t+g(x_1,x_2)\mathrm{d}B_t,
\end{align}
 where $f(y^*,0,0)=g(y^*,0)=0$ and
 \begin{align*}
 \|\frac{\partial f}{\partial x_{i}}\|\leq L_{i},~
\|\frac{\partial g}{\partial x_{i}}\|\leq N_{i};~~\frac{\partial f}{\partial u}\geq I_n,~\forall x_1,x_2,u,
\end{align*}
then for any positive quadruple $(L_1,L_2,N_1,N_2)$, the uncertain stochastic system (\ref{nou}) is globally stabilizable by the PD control (\ref{pdd}), see Theorem 3.9 in \cite{zhang2021}.  Moreover, the selection of the PD parameters $k_p$ and $k_d$ has wide flexibility, since they can be arbitrarily chosen from an open and unbounded set in $\mathbb{R}^2$. Thus, one might naturally conjecture that this result can be extended to the more general system (1) considered in this article, where $g(\cdot)$ is a function of both the state variables and the control variables. To be precisely, for any given positive constants $L_1,L_2,N_1,N_2$ and $M$,  an open and unbounded PD parameter set could be constructed, from which the PD control (2) has the ability to globally stabilize the system (1),  for all functions  $f$ and $g$ that satisfy Assumptions 1 and 2.
Surprisingly, the answer of the above problem is no. In fact, these five constants have to meet suitable constraints before such stabilizing PD parameters can be found.
\end{Remark}
\section{Main Results}
\subsection{Uncertain Nonlinear Stochastic System}
For given positive constants $L_1,L_2,N_1$ and $N_2$, we first define a family of parameter set $\{\Omega_0(M),~M\geq 0\}$ as follows:
\begin{align}\label{gain}
 \Omega_0:=\left\{(k_p,k_d)\in\mathbb{R}_+^2\left|
\begin{array}{c}
		k_p^2>\bar k+k_dT_1^2\\
		k_d^2-k_p>\bar k+k_dT_2^2
\end{array}
\right.\!\!\!
\right\},
\end{align}
where $\bar k:=(L_1+L_2)(k_p+k_d)$, and $T_1$, $T_2$ are defined by
\begin{align}\label{1222}
T_1:=&N_1+Mk_p,~~ T_2:=N_2+Mk_d.
\end{align}
Next, we list some geometric properties of the set $\Omega_0$:
\begin{itemize}
\item If $M=0$, $\Omega_0$ is an open and unbounded set in $\mathbb{R}^2$;
\item The range of $\Omega_{0}$ will shrink as $M$ increase, i.e. \begin{align*}
\Omega_0(M_1)\subset \Omega_0(M_2),~~\text{if}~0\le M_2< M_1;
\end{align*}
\item  $\Omega_0=\emptyset$ if $M\geq M_0^*$, where $M_0^*$ is the unique positive solution of
$16L_1s^4+16N_1s^3+4L_2s^2+4N_2s=1.$
\end{itemize}

Let  $M_{1}^*$ be the supremum of the set consisting of $M$ that makes $\Omega_0$ nonempty. More precisely,
\begin{align*}
M_{1}^*:=\sup \big\{M>0: \Omega_0(M)\neq \emptyset\big\}.
\end{align*}

\begin{Theorem}Consider the nonlinear stochastic system (1)-(2), where Assumptions 1-2 are satisfied.

(i) If $0\le M<M_1^*$, system (1) is stabilizable by the PD control (2).  Moreover, the stabilizing PD parameters can be selected from $\Omega_0$.
%Moreover, (\ref{3.0}) is satisfied  for all $f(\cdot)\in\mathcal{F}_{L_{1},L_{2}}$ and  $g(\cdot)\in \mathcal{G}_{N_{1},N_{2},M}$, provided that $(k_p,k_d)\in \Omega$.

(ii) If $M\geq M_2^*$, where $M_2^*$ is the unique positive root of the following quartic polynomial:
\begin{align}\label{m*}4L_1s^4+4N_1s^3+2L_2s^2+2N_2s- 1=0,\end{align}
 system (1) is not stabilizable by the PD control (2).
\end{Theorem}
The proof of Theorem 1 will be provided in the next section.

%then for any given PD parameters $(k_p,k_d)\in\mathbb{R}^2$, there always exist some $f(\cdot)\in\mathcal{F}_{L_{1},L_{2}}$ and $g(\cdot)\in \mathcal{G}_{N_{1},N_{2},M}$, such that $\mathbb{E} \left[\|e(t)\|^2+\|x_2(t)\|^2\right]$ will not converge to zero for some initial state.

\begin{Remark}
Note that $M\geq M_2^*$ is equivalent to
\begin{align}U:=4L_1M^4+4N_1M^3+2L_2M^2+2N_2M\geq 1.\end{align}
Hence, it can be seen from Theorem 1(ii) that, if we regard the quantity $U$ as a \emph{measure} of system uncertainty, the PD control (2) will have fundamental limitations in dealing with the uncertain nonlinear stochastic system (1), once the uncertainty of the system is too large, namely, $U\geq 1$.
\end{Remark}

\begin{Remark} The constant $M_1^*>0$ and satisfies
\begin{align}
16L_1M_{1}^{*4}+\!16N_1M_{1}^{*3}+\!4L_2M_{1}^{*2}+\!4N_2M_{1}^*\le\! 1.
\end{align}
Therefore,  $M_1^*<M_2^*$. So, one naturally ask, whether system (1) is stabilizable if $M_1^*\le M<M_2^*$? Further, does it exist a positive constant $M^*$, such that system (1) is stabilizable by the PD control (2) if and only if $M<M^*$?
In general, these problems can be very challenging due to the inherent nonlinearity, uncertainties and the strong coupling of high-dimensional state variables and non-affine control input, and remains open. However,  the stabilizability problems have been solved in this article, when system (1) has a specific linear structure. In fact, the necessary and sufficient condition for a class of uncertain linear stochastic systems to be stabilized by the PD control (2) is $U<1$,  see Theorem 2 for details.
\end{Remark}

\subsection{Uncertain Linear Stochastic System}
%If the uncertain stochastic system (1) has a linear structure, the results in Theorem 1 can be improved.
Consider the following uncertain linear stochastic system:
\begin{align}\label{sys}
\mathrm{d} x_1=&x_2 \mathrm{d}t\nonumber\\
\mathrm{d}x_2=&(ax_1+bx_2+u)\mathrm{d}t+(cx_1+dx_2+eu)\mathrm{d}B_t,
\end{align}
where $x_1\in\mathbb{R}^n,~u\in\mathbb{R}^n,$ and $a$, $b$, $c$, $d$ and $e$ are unknown $n\times n$ constant matrices with known upper bounds, namely,
\begin{align}\label{ab}
\|a\|\le L_1, ~\!\!\|b\|\le L_2,~\! \!\|c\|\le N_1, ~\!\!\|d\|\le N_2,~\!\!\|e\|\le M,
\end{align}
where $L_1, L_2, N_1, N_2$ are positive constants, and $M$ is nonnegative.
Without loss of generality, assume that $y^*=0$, then the  PD control (2) takes the following form:
\begin{align}\label{14}u(t)=-k_p x_1(t)-k_d x_2(t).\end{align}

Next, we will present a necessary and sufficient condition on the five constants, under which the uncertain stochastic system (\ref{sys}) is stabilizable by the PD control (\ref{14}). Moreover, necessary and sufficient conditions for the choice of PD parameters are also provided.

\begin{Theorem}Consider the system (\ref{sys}), where (\ref{ab}) is satisfied. Then, the necessary and sufficient condition for system (\ref{sys}) to be stabilizable by the PD control (\ref{14}) is
\begin{align}
\label{M11}4L_1M^4+4N_1M^3+2L_2M^2+2N_2M<1.
\end{align}
Moreover, under (\ref{M11}), the closed-loop system will satisfy (3)
%\begin{align}\label{141}
%\lim_{t\to\infty} \mathbb{E} \left[\|x_1(t)\|^2\!+\!\|x_2(t)\|^2\right]=0,~\forall (x_1(0),x_2(0)),\end{align}
for all constant matrices $a$, $b$, $c$, $d$ and $e$ satisfying (\ref{ab}), if and only if $k_p$ and $k_d$ are chosen from the following set:
\begin{align}\label{pD}
\Omega:=\left\{(k_p,k_d)~\!\left|~\!k_p>L_1,~\!2\bar{k}_1\bar{k}_2>T_1^2+\bar{k}_1 T_2^2\right.\right\},
\end{align}
where  $T_1$, $T_2$ are defined in (\ref{1222}) and $\bar{k}_1$, $\bar{k}_2$ are defined by
 \begin{align}\label{12}
&~\bar{k}_1:=k_p-L_1,~~~~~ \bar{k}_2:=k_d-L_2.\end{align}
%where $\bar{k}_1, \bar{k}_2, T_1$ and $T_2$ are defined in (\ref{12})$-$(\ref{1222}).
\end{Theorem}
\vskip 0.25cm
\begin{Remark}We provide some geometric properties of the parameter set $\Omega$ defined by (\ref{pD}):
\begin{itemize}
\item The set $\Omega_0$ defined in (\ref{gain}) is a subset of $\Omega$;
\item If $M=0$, $\Omega$ is an open and unbounded subset in $\mathbb{R}^2$ for any  positive constants $L_1$, $L_2$, $N_1$ and $N_2$;
\item  If $M>0$ and (\ref{M11}) holds, $\Omega$ is an open and \emph{bounded convex} subset in $\mathbb{R}^2$.\end{itemize}
Hence, the controller gains $k_p$ and $k_d$ cannot be chosen sufficiently large for the case $M>0$, which is a significant difference from the existing literature on PD or PID controlled nonlinear uncertain systems, see e.g. \cite{Khalil},\cite{zhao2017},\cite{cong2017},\cite{zhang2021}.
\end{Remark}

%Figure 1 depicts the shape of $\Omega$ under different values of $M$, with $(L_1,L_2,N_1,N_2)=(1/4,1/2,1,1/2)$. It can be seen intuitively that the set $\Omega$ is reflectional symmetric, and $\Omega$ will becomes smaller, when $M$ becomes larger.
%\begin{figure}
%\includegraphics[width=9cm]{pdset}
%\caption{The Shape of Set $\Omega$ under Different Values of $M$}
%\end{figure}
\section{Proofs of the Main Results}
\subsection{Proof of Theorem 1}

First, we prove the first half of Theorem 1 in three steps.

\emph{Step 1: (Some properties of the PD parameter set $\Omega_0$)}

Firstly, for given positive constants $L_1,L_2,N_1,N_2$ and $M\geq 0$,
we define a set $\Omega'$ as follows:
\begin{align}\label{omega'}
\Omega':=\left\{(k_p,k_d)~\!\left|~k_p>L_1,~\bar{k}_1\bar{k}_2>T_1^2+\bar{k}_1 T_2^2\right.\right\},
\end{align}
where $T_1$, $T_2$, $\bar{k}_1$ and $\bar{k}_2$ are defined in (\ref{1222}) and  (\ref{12}).

\emph{Property 1:} The sets $\Omega_0$, $\Omega'$ and $\Omega$ satisfy
\begin{align}\label{subset1}
\Omega_0\subset \Omega'\subset \Omega,
\end{align}
where $\Omega_0$ and $\Omega$ are defined in  (\ref{gain}) and (\ref{pD}), respectively.

The inclusion $\Omega'\subset \Omega$ is obvious by the definitions of $\Omega'$ and $\Omega$. We only need to show $\Omega_0\subset \Omega'$. Indeed, if $(k_p,k_d)\in \Omega_0$, then by definition (\ref{gain}), we know
\begin{align}
k_p^2>\bar k:=(L_1+L_2)(k_p+k_d)>k_p (L_1+L_2),
\end{align}
which yields $k_p>L_1$. Moreover, combine
$k_p^2>\bar k+k_dT_1^2$ with $\bar k>k_pL_1$, it can be obtained that
$k_p^2-k_pL_1>k_dT_1^2$, hence $k_p-L_1>k_dT_1^2/k_p$. Recall $\bar k_1:=k_p-L_1$, we have
\begin{align}\label{18}
\bar k_1>k_dT_1^2/k_p.
\end{align}
On the other hand, since $k_d^2-k_p>\bar k+k_dT_2^2>k_dL_2+k_dT_2^2$, we have
$k_d(k_d-L_2)>k_p+k_dT_2^2$. Therefore, we have
\begin{align} \label{19}
\bar k_2:=k_d-L_2>k_p/k_d+T_2^2.
\end{align}
Combine (\ref{18}) with (\ref{19}), it is easy to obtain
\begin{align}\label{11}
\bar k_1\left(\bar k_2-T_2^2\right)>T_1^2.
\end{align}
From (\ref{11}) and recall $k_p>L_1$, we conclude that $(k_p,k_d)\in \Omega'$.

\emph{Property 2:}
 $\Omega_0$ will become smaller when $M$ increase, i.e.
\begin{align}\label{subset}
\Omega_0(M_1)\subset \Omega_0(M_2),~~\text{if}~0\le M_2< M_1.
\end{align}
In fact, let $0\le M_2< M_1$, and suppose $(k_p,k_d)\in  \Omega_0(M_1)$, then by (\ref{gain}) and note that $k_p>0,k_d>0$, it is easy to obtain
\begin{align*}
&k_p^2>\bar k+k_d(N_1+M_1k_p)^2>\bar k+k_d(N_1+M_2k_p)^2,\\
&k_d^2-k_p>\bar k+k_d(N_2+M_1k_d)^2>\bar k+k_d(N_2+M_2k_d)^2,
\end{align*}
hence, $(k_p,k_d)\in \Omega_0(M_2)$, which yields the relationship (\ref{subset}).

\emph{Property 3:} If $M=0$, $\Omega_0$ is an open and unbounded set.

To this end,  let $k_{p}=k_{d}=k>0$, then $$k_p^2-\bar k-k_dT_1^2=k^2-2k(L_1+L_2)-kN_1^2>0, ~\text{as}~k\to \infty.$$ Similarly, $k_d^2-k_p-\bar k-k_dT_2^2>0$ for $k$ large enough. Consequently, $\Omega_0$ is  open and unbounded when $M=0$.

\emph{Property 4:} For given positive constants $L_1$, $L_2$, $N_1$ and $N_2$,   let $M_0^*$ be the unique positive solution of the polynomial
\begin{align}
16L_1s^4+16N_1s^3+4L_2s^2+4N_2s=1,
\end{align}
then $\Omega_0=\emptyset$ if $M\geq M_0^*$.

First, by the definition of $M_0^*$, we know that $M\geq M_0^*$ is equivalent to $16L_1M^4+16N_1M^3+4L_2M^2+4N_2M\geq 1$. From Lemma 2 in Appendix A, we know  $\Omega'$ is empty if and only if $M\geq M_0^*$. Besides, by the inclusion relationship (\ref{subset1}), we conclude that $\Omega_0$ is also empty if $M\geq M_0^*$.

Let $M_{1}^*$ be the supremum that makes $\Omega_0$ nonempty, i.e.,
\begin{align}\label{m1}
M_{1}^*:=\sup \big\{M>0: \Omega_0(M)\neq \emptyset\big\}.
\end{align}
Then, it is easy to obtain the following facts:
\begin{itemize}
\item The set $\Omega_0$ is not empty, if $0\le M<M_{1}^*$;
\item The constant $M_{1}^*$ depends on $L_1,L_2,N_1,N_2$ only;
\item  $M_{1}^*\le M_0^*$, i.e.,
$16L_1M_{1}^{*4}+\!16N_1M_{1}^{*3}+\!4L_2M_{1}^{*2}+\!4N_2M_{1}^*\le\! 1.$
\end{itemize}

\emph{Step 2: (Write the closed-loop system into a linearity-like form)}
~Let us denote
\begin{align}
&z_1(t):=-e(t)=x_1(t)-y^*,~~z_2(t):=-\dot e(t)=x_2(t),
\end{align}
then  the PD control (2) can be rewritten as  $u(t)=-k_p z_1(t)-k_d z_2(t),$ and the PD controlled system (1)-(2) turns into
\begin{align}\label{g000}
\mathrm{d} z_1&=~z_2 \mathrm{d}t\nonumber\\
\mathrm{d}z_2&=~f(z_1+y^*,z_2,u)\mathrm{d}t+ g(z_1+y^*,z_2,u)\mathrm{d}B_t.\\
~~\!u&=~-k_p z_1-k_d z_2\nonumber
\end{align}
By Assumption 1, we know that $(z_1,z_2)=(0,0)\in\mathbb{R}^{2n}$ is an equilibrium of (\ref{g000}). Besides,  recall $f(y^*,0,0)=0$ and note that $f\in \mathcal{F}_{L_{1},L_{2}}$,
 one can obtain(details can be found in \cite{zhao2020}):
\begin{align}\label{f}
 f(z_1\!+\!y^*,z_2,u)\!=\!a(z_1) z_1\!+\!b(z_1,z_2)z_2\!+\!\theta(z_1,z_2,u) u,
\end{align}
where $a$, $b$ and $\theta$ are $n\times n$ matrices satisfying
 \begin{equation}\label{gg1}
 \|a(z_1)\|\le L_1,~\|b(z_1,z_2)\|\le L_2,~\theta(z_1,z_2,u)\geq I_n,
 \end{equation}
for all $z_1,z_2,u$. Similarly, since $g(y^*,0,0)=0$ and $g\in \mathcal{G}_{N_{1},N_{2},M}$, the function $g(z_1\!+\!y^*,z_2,u)$ can be expressed by
 \begin{align}\label{g}
 g(z_1\!+\!y^*,z_2,u)\!=\!c(z_1) z_1\!+\!d(z_1,z_2)z_2\!+\!e(z_1,z_2,u) u,
\end{align}  where
  $c$, $d$ and $e$ are $n\times n$ matrices satisfying
 \begin{equation}\label{gg2}
 \|c(z_1)\|\le N_1,~\|d(z_1,z_2)\|\le N_2,~\|e(z_1,z_2,u)\|\le M.
 \end{equation}
By the expressions of $f$ and $g$ in (\ref{f}) and (\ref{g}), the nonlinear  system (\ref{g000}) turns into the linearity-like form:
\begin{align}\label{g111}
\mathrm{d} {z}_1=&z_2\mathrm{d}t\nonumber\\
\mathrm{d} {z}_2=&\big[\hat a z_1+\hat b z_2\big]\mathrm{d}t+\big[\hat cz_1+\hat dz_2\big]\mathrm{d}B_t,
\end{align}
where $\hat a$, $\hat b$, $\hat c$ and $\hat d$ are nonlinear (matrix-valued) functions of $z=(z_1,z_2)$ defined by
\begin{align}\hat a=a(z_1)-k_p\theta(z,u),
~~ \hat b=b(z_1,z_2)-k_d\theta(z,u),\label{26}\\
\hat c=c(z_1)-k_pe(z,u),
~~\hat d=d(z_1,z_2)-k_de(z,u),\label{27}\end{align}
with $u=k_pz_1+k_d z_2.$
%Denote
% $z=\begin{bmatrix}
% z_1\\z_2\end{bmatrix}\in\mathbb{R}^{2n}$,
% and define two $2n\times 2n$ matrices
%\begin{align}A(z):=
%\begin{bmatrix}0_n&I_n\\\widehat a&\widehat b\end{bmatrix},~~
%B(z)
%:=
%\begin{bmatrix}0_n&0_n\\\widehat c&~\widehat d
%\end{bmatrix},\end{align}
%then the nonlinear system (\ref{g111}) can be rewritten in a  linearity-like form:
%\begin{align}\label{s2}
%\mathrm{d}z= A(z)z\mathrm{d}t+B(z)z\mathrm{d}B_t,~~ z\in\mathbb{R}^{2n}.
%\end{align}

Now, suppose that $M<M_{1}^*$ and $(k_p,k_d)\in \Omega_0$. In the next two steps, we proceed to prove the closed-loop system (1)-(2) will satisfy the performance (3).

\emph{Step 3: (Construction of  Lyapunov function)}

We adopt a similar Lyapunov function $V(z)$ as that used for deterministic system(see \cite{zhao2020}),
\begin{equation}
V(z)=k_pk_dz_1^{\mathsf{T}}z_1+k_pz_1^{\mathsf{T}}z_2+k_dz_2^{\mathsf{T}}z_2/2,~\!~z=(z_1,z_2).\label{eq:v1}
\end{equation}
%where $P$ is a $2n\times 2n$ symmetric matrix defined by
%	\[
%	P=\begin{bmatrix}\begin{array}{cc}
%			2k_{p}k_{d}I_{n} & \quad k_{p}I_{n}\\
%			k_{p}I_{n} & \quad k_{d}I_{n}
%	\end{array}\end{bmatrix},
%	\]
%and $I_n$ is the $n\times n$ identity matrix.	
Note that $
V=\|\sqrt{k_pk_d}z_1+\frac{1}{2}\sqrt{k_p/k_d}z_2\|^2+\frac{1}{2}(k_d-\frac{k_p}{2k_d})\|z_2\|^2
$
and that $k_d^2>k_p$, we know that $V(z)$ is positive definite.

%$$P=\begin{bmatrix}2k_pk_d\underline b&k_p\\
%k_p&k_d\end{bmatrix}\begin{bmatrix}0&-I\\ k_p\theta-a&k_d\theta-b\end{bmatrix}$$

\emph{Step 4: (Stability analysis based on Lyapunov methods)}

By some manipulations, the operator $L$ acting on the function $V(z)$ along the trajectories of (\ref{g111}) is given by
\begin{align}\label{LV}
LV(z)=\underbrace{\frac{\partial V}{\partial z}
\begin{bmatrix}z_1\\
\hat a z_1\!+\hat b z_2
\end{bmatrix}}_{\mathrm{I}}\!+\underbrace{\frac{k_d}{2}\left\|\hat c z_1\!+\!\hat d z_2\right\|^2}_{\mathrm{II}}.
\end{align}
Denote $A=\begin{bmatrix}0_n&I_n\\\hat a&\hat b\end{bmatrix}$ and $P=\begin{bmatrix}
			2k_{p}k_{d} I_n &  k_{p} I_n\\
			k_{p} I_n & k_{d} I_n
	\end{bmatrix}$, where $0_n$ is the $n\times n$ zero matrix,
then the first term can be estimated as follows(see the proof of Proposition 4.3 in \cite{zhao2020}):
\begin{align}\label{I}
\mathrm{I}=&z^{\mathsf{T}}\left(PA+A^{\mathsf{T}}P\right)z\nonumber\\
\le& -(k_p^2-\bar k)\|z_1\|^2-(k_d^2-k_p-\bar k)\|z_2\|^2.
\end{align}
By the definitions of $\hat c$, $\hat d$ in (\ref{27}) and the properties (\ref{gg2}), we know that
$\|\hat c\|\le N_1+k_pM=T_1$ and $\|\hat d\|\le N_2+k_dM=T_2$. Therefore,
the second term has the following upper bound:
\begin{align}\label{II}
\mathrm{II}=\frac{k_d}{2}\big\|\hat c z_1\!+\!\hat d z_2\big\|^2\le k_d\left(T_1^2\|z_1\|^2+T_2^2\|z_2\|^2\right).
\end{align}
Combine (\ref{I}) and (\ref{II}), we obtain the upper bounds of $LV(z)$:
	\begin{align*}
		LV(z)\leq\! -(k_p^2\!-\!\bar k\!-\!k_dT_1^2)\|z_1\|^2-(k_d^2\!-\!k_p\!-\!\bar k\!-\!k_dT_2^2)\|z_2\|^2.
	\end{align*}
Since $(k_p,k_d)\in \Omega_0$, we know that
	\begin{equation}
		LV(z)\leq-\eta \|z\|^2, ~~\forall z\in\mathbb{R}^{2n}.\label{eq:lv1}
	\end{equation}
for some positive constant $\eta$. Recall $z_1(t)=-e(t),~z_2(t)=-\dot e(t)$, we conclude that
the PD control system (1)-(2) will satisfy the control objective (3).

Next, we prove the second half of Theorem 1. For this, it suffices to show the following statement:

If there exist some $(k_p,k_d)\in\mathbb{R}^2$,  such that the closed-loop system (1)-(2) satisfies the performance $(3)$ for all $f(\cdot)$ and $g(\cdot)$ satisfying Assumptions 1-2, then $M<M_2^*$.

By the definition of $M_2^*$ in (\ref{m*}), one can see that
$$M< M_2^*\Longleftrightarrow4L_1M^4+4N_1M^3+2L_2M^2+2N_2M<1.$$
Now, suppose that the functions $f(\cdot)$ and $g(\cdot)$ are given by
 \begin{align}\label{f0}
 &f(x_1,x_2,u)=a(x_1-y^*)+b x_2+u,~x_1,x_2,u\in\mathbb{R}^n,\\\label{g0}
 & g(x_1,x_2,u)=c(x_1-y^*)\!+d x_2-eu,~x_1,x_2,u\in\mathbb{R}^n,\end{align}
where $a,b,c,d,e$ are five real numbers satisfying $|a|\le L_1$, $|b|\le L_2$, $|c|\le N_1$, $|d|\le N_2$ and $|e|\le M$.  Then it is easy to see that Assumptions 1 and 2 hold.

Denote $z_1(t)=x_1(t)-y^*$ and $z_2(t)=x_2(t)$,
%\begin{align}\label{30ee}
%\mathrm{d} z_1=&z_2 \mathrm{d}t\nonumber\\
%\mathrm{d} z_2=&\big[a_0z_1+b_0z_2\big]\mathrm{d}t+\big[c_0z_1+d_0z_2\big]\mathrm{d}B_t,
%\end{align}
 $$z(t):=\begin{bmatrix}z_1(t)\\z_2(t)\end{bmatrix},~A:=\begin{bmatrix}0_n&I_n\\a_0I_n&b_0I_n\end{bmatrix},~
B:=\begin{bmatrix}0_n&0_n\\c_0I_n&d_0I_n\end{bmatrix},$$
where
\begin{align}\label{a0d0}
a_0=a-k_p, b_0=b-k_d, c_0=c+k_p e,d_0=d+k_de,
\end{align}
then closed-loop equation (1)-(2) with $f$ and $g$ defined by (\ref{f0}) and (\ref{g0}) turns into:
\begin{align}\label{30eq1}
\mathrm{d}z=Az \mathrm{d}t+B z\mathrm{d}B_t.
\end{align}
Define a $2n\times 2n$ time-varying matrix $P(t):=\begin{bmatrix}
p(t)&r(t)\\r^{\mathsf{T}}(t)&q(t)
\end{bmatrix},$
where $p(t)$, $r(t)$ and $q(t)$ are $n\times n$ matrix defined by
\begin{align}
p(t):=&~\mathbb{E}\left[z_1(t)z_1^{\mathsf{T}}(t)\right], ~r(t):=\mathbb{E}\left[z_1(t)z_2^{\mathsf{T}}(t)\right],\\
q(t):=&~\mathbb{E}\left[z_2(t)z_2^{\mathsf{T}}(t)\right],\end{align}
then it can be seen that $P(t)=\mathbb{E}\left[ z(t)z^{\mathsf{T}}(t)\right]$.  From Theorem 8.5.5 in \cite{Arnold}, we know that $P(t)$ is the unique nonnegative-definite symmetric solution of the equation
\begin{align}\label{34}
\frac{\mathrm{d} P}{\mathrm{d} t}=AP(t)+P(t)A^{\mathsf{T}}+BP(t)B^{\mathsf{T}}.
\end{align}
From (\ref{34}), it can be obtained that
\begin{align}
&\dot p=r+r^{\mathsf{T}}\nonumber\\
&\dot r=a_0p+b_0r+q\label{68}\\
&\dot q=c_0^2p+(a_0+c_0d_0)\left(r+r^{\mathsf{T}}\right)+\left(2b_0+d_0^2\right)q\nonumber.
\end{align}
Let $r_0(t)=(r(t)+r^{\mathsf{T}}(t))/2$ and define
\begin{align}\label{q}
Q=\begin{bmatrix}
0&2&0\\
a_0&b_0&1\\
c_0^2&2(a_0+c_0d_0)&2b_0+d_0^2
\end{bmatrix},
\end{align}
then  it follows from (\ref{68}) that
\begin{align}
\frac{\mathrm{d}}{\mathrm{d}t}\begin{bmatrix}p(t)\\r_0(t)\\q(t)\end{bmatrix}
=Q\otimes I_n\begin{bmatrix}p(t)\\r_0(t)\\q(t)\end{bmatrix},
\end{align}
where $\otimes$ denotes the  Kronecker product.

Since for any initial state $(z_1(0), z_2(0))\in\mathbb{R}^{2n}$, the solution of (\ref{30eq1}) satisfies  $\lim_{t\to\infty}\mathbb{E} \left[\|z_1(t)\|^2+\|z_2(t)\|^2\right]=0$, which implies that $\lim_{t\to\infty}\|P(t)\|=0$ for all initial state $(z_1(0), z_2(0))$.
We conclude that $Q\otimes I_n$ is a Hurwitz matrix. Note that the matrix $Q\otimes I_n$ shares the same spectrum with $Q$. Hence,  $Q$ is also Hurwitz.

%Next, we aim to find  equivalent conditions (on $L_1$, $L_2$, $N_1$, $N_2$ and $M$) to ensure that there exist $k_p$ and $k_d$, such that the closed-loop control system (3) is exponentially stable in mean square,  for all $|a|\le L_1$, $|b|\le L_2$, $|c|\le N_1$, $|d|\le N_2$ and $|e|\le M$.

%Next, we need to find conditions to ensure the matrix $Q$ to be Hurwitz.
From the expression of $Q$ in (\ref{q}),  the characteristic polynomial of $Q$ can be calculated as follows:
\begin{align}
\det (\lambda I_3-Q)=\lambda^3+\alpha_2 \lambda^2+\alpha_1\lambda+\alpha_0,
\end{align}
where  $\alpha_0$, $\alpha_1$ and $\alpha_2$ are given by
\begin{align}
\alpha_1=&b_0d_0^2+2b_0^2-4a_0-2c_0d_0 \\
\alpha_0=&2(2a_0b_0+a_0d_0^2-c_0^2),~\alpha_2=-(3b_0+d_0^2).\label{a2}\end{align}
From the Routh-Hurwitz stability criterion for third order polynomials, the matrix $Q$ is Hurwitz if and only if the following inequalities holds:
\begin{align}\label{42}
\alpha_2>0,~~~ \alpha_0>0,~~~\alpha_1\alpha_2>\alpha_0.
\end{align}

We next proceed to show the following statement:

Suppose that the matrix $Q$ defined in (\ref{q}) is Hurwitz for all  $|a|\le L_1$, $|b|\le L_2$, $|c|\le N_1$, $|d|\le N_2$, $|e|\le M$, then the parameters $k_p$ and $k_d$ belong to the set $\Omega$ defined in (\ref{pD}).

Proof. First, from the definitions of $\alpha_2$ and $b_0$ in (\ref{a2}) and (\ref{a0d0}), we have $\alpha_2=-(3b_0+d_0^2)=3(k_d-b)-d_0^2$. In addition, since $\alpha_2>0$, it follows that $3(k_d-b)\geq \alpha_2>0$.
Choose $b=L_2,$ we conclude that
\begin{align}\label{.1}k_d> L_2.\end{align}
Next, suppose for all $|a|\le L_1, |b|\le L_2, |c|\le N_1, |d|\le N_2, |e|\le M$, there is $$\alpha_0=2(a-k_p)(b-k_d)+(a-k_p)(d-ek_d)^2-(c-ek_p)^2>0,$$
Choose $c=d=e=0$, it follows from (\ref{.1})  that $k_p>L_1$.
Moreover, if we choose $a=L_1$, $b=L_2$, $c=-N_1$, $d=-N_2$ and $e=M$, then we have
\begin{align}\label{2}
2\bar k_1\bar k_2-\bar k_1(N_2+Mk_d)^2-(N_1+Mk_p)^2>0.
\end{align}
Combine (\ref{.1})-(\ref{2}), we conclude that $(k_p,k_d)$ belongs to $\Omega$.

Finally,  by Lemma 1 in Appendix A, we know that the necessary and sufficient condition for the set $\Omega$ to be non-empty is $4L_1M^4+4N_1M^3+2L_2M^2+2N_2M<1.$ Hence, if $M\geq M_2^*$, where $M_2^*$ is the unique positive solution of the equation (\ref{m*}), then $\Omega$ is empty, and thus there does not exist  $k_p$ and $k_d$ such that $Q$ is Hurwitz for all  $|a|\le L_1$, $|b|\le L_2$, $|c|\le N_1$, $|d|\le N_2$ and $|e|\le M$. Therefore, the system (1) is not stabilizable by the PD control (2).  \hfill$\square$

\subsection{Proof of Theorem 2}
\emph{Sufficiency:}
Suppose that (\ref{M11}) is satisfied. From Lemma 1 in Appendix A, we know that the set $\Omega$ defined by (\ref{pD}) is not empty. Now, suppose $(k_p,k_d)\in \Omega$ and the matrices $a$, $b$, $c$, $d$ and $e$ satisfy (\ref{ab}), we proceed to show that the closed-loop system (\ref{sys}) and (\ref{14})  will satisfy $\lim_{t\to\infty} \mathbb{E} \left[\|x_1(t)\|^2\!+\!\|x_2(t)\|^2\right]=0$ for all initial states.
Substituting (\ref{14}) into (\ref{sys}), we have
\begin{align}\label{close}
\mathrm{d} x_1=&x_2 \mathrm{d}t\nonumber\\
\mathrm{d} x_2=&\big[a_0x_1+b_0x_2\big]\mathrm{d}t+\big[c_0x_1+d_0x_2\big]\mathrm{d}B_t,
\end{align}
where $a_0$, $b_0$, $c_0$, $d_0$ are $n\times n$ constant matrices defined by
\begin{align}
a_0\!=\!a\!-\!k_pI_n, b_0\!=\!b\!-\!k_dI_n, c_0\!=\!c\!-\!k_pe, d_0\!=\!d\!-\!k_de.
\end{align}
Define
\begin{align}
A:=\begin{bmatrix}0_n&I_n\\a_0&b_0\end{bmatrix},~~X:=\begin{bmatrix}x_1\\x_2\end{bmatrix},
~~B:=\begin{bmatrix}0_n&0_n\\c_0&d_0\end{bmatrix},
\end{align}
then (\ref{close}) can be rewritten in a more compact form:
\begin{align}\label{30eq}
\mathrm{d}X=AX \mathrm{d}t+B X\mathrm{d}B_t.
\end{align}
Let us define a $2n\times 2n$ matrix $P$ as follows:
\begin{align}\label{51}P:=\begin{bmatrix}
m&rI_n\\
rI_n&I_n
\end{bmatrix}, \end{align}
where $m$ is an $n\times n$ matrix defined by
\begin{align}\label{m}
m:=-rb_0-a_0^{\mathsf{T}}-c_0^{\mathsf{T}} d_0,
\end{align}
and $r>0$ is a constant defined by
\begin{align}r:=\left(2\bar k_1\bar k_2+T_1^2-\bar k_1 T_2^2\right)/\left(4\bar k_1\right).\end{align}
Now, let $V(X)=X^{\mathsf{T}} PX$, where $P$ is defined in (\ref{51}),
then  the differential
operator $L$ acting on $V$ is
\begin{align*}
&LV(X)=X^{\mathsf{T}} \left(PA+A^{\mathsf{T}}P+B^{\mathsf{T}}PB\right)X\\
=&X^{\mathsf{T}}\begin{bmatrix}2ra_0^{\text{sym}}+c_0^{\mathsf{T}}c_0&m+rb_0+a_0^{\mathsf{T}}+c_0^{\mathsf{T}}d_0\\
m^{\mathsf{T}}+rb_0^{\mathsf{T}}+a_0+d_0^{\mathsf{T}}c_0&2rI_n+2b_0^{\text{sym}}+d_0^{\mathsf{T}}d_0\end{bmatrix}X,
%\\
%=&X^{\mathsf{T}}\begin{bmatrix}~~2ra_0^{\text{sym}}+c_0^{\mathsf{T}}c_0&0_n\\
%0_n&2r+2b_0^{\text{sym}}+d_0^{\mathsf{T}}d_0\end{bmatrix}X,
\end{align*}
where $$a_0^{\text{sym}}=(a_0+a_0^{\mathsf{T}})/2, ~b_0^{\text{sym}}=(b_0+b_0^{\mathsf{T}})/2$$ are the symmetrization matrices of $a_0$ and $b_0$.
Denote $Q=PA+A^{\mathsf{T}}P+B^{\mathsf{T}}PB$, then from (\ref{m}), it is easy to obtain
\begin{align}
Q=\begin{bmatrix}~~2ra_0^{\text{sym}}+c_0^{\mathsf{T}}c_0&0_n\\
0_n&2r+2b_0^{\text{sym}}+d_0^{\mathsf{T}}d_0\end{bmatrix}.
\end{align}
Note that $\|a\|\le L_1$ and  $\|b\|\le L_2$, it follows that  \begin{align}\label{55}\lambda_{\max}\left[a_0^{\text{sym}}\right]\le L_1-k_p=-\bar k_1,
~~\lambda_{\max}\left[ b_0^{\text{sym}}\right]\le -\bar k_2.\end{align}
Moreover, since $\|c\|\le N_1$,  $\|d\|\le N_2$, $\|e\|\le M$, we have
$$\|c_0\|=\|c-k_pe\|\le T_1,~ \|d_0\|=\|d-k_de\|\le T_2,$$
where $T_1$, $T_2$ are defined in (\ref{1222}).
From (\ref{55}), and recall  $(k_p,k_d)\in\Omega$, it can be seen that
\begin{align}
&\lambda_{\max}\left[2ra_0^{\text{sym}}+c_0^{\mathsf{T}}c_0\right]
\le -2\bar k_1r+T_1^2\nonumber\\
=&\left(\bar k_1T_2^2+T_1^2-2\bar k_1\bar k_2\right)/2<0,\\
&\lambda_{\max}\left[2r+2b_0^{\text{sym}}+d_0^{\mathsf{T}}d_0\right]<2r-2\bar k_2+T_2^2\nonumber\\
=&\left(\bar k_1T_2^2+T_1^2-2\bar k_1\bar k_2\right)/(2\bar k_1)<0,
\end{align}
which implies $LV(X)$ is a negative definite function.

Finally, we prove $V(X)=X^{\mathsf{T}} PX$ is positive definite. By the definition of $P$ in (\ref{51}), it suffices to show
$m^{\text{sym}}-r^2I_n>0$. Note that
$m^{\text{sym}}=-r {b}_0^{\text{sym}}-{a}_0^{\text{sym}}-[c_0^{\mathsf{T}}d_0]^{\text{sym}},$
we have
$$\lambda_{\text{min}}\big[ m^{\text{sym}}\big]\geq
 -r\lambda_{\text{max}}\big[{b}_0^{\text{sym}}\big]-\lambda_{\text{max}}\big[{a}_0^{\text{sym}}\big]-
\left\| [c_0^{\mathsf{T}}d_0]^{\text{sym}}\right\|.$$
Therefore, it follows from (\ref{55}) that
\begin{align}
&\lambda_{\text{min}}\big[ m^{\text{sym}}-r^2I_n\big]~\geq ~r\bar k_2-r^2+\bar k_1-T_1T_2\nonumber\\
=&-(r-\bar k_2/2)^2 +\bar k_2^2/4+\bar k_1-T_1T_2\nonumber\\
=&-\big(T_1^2/(4\bar k_1)-T_2^2/4\big)^2 +\bar k_2^2/4+\bar k_1-T_1T_2.
\end{align}
Consequently,
\begin{align}
&\lambda_{\text{min}}\left[16\bar k_1^2( m^{\text{sym}}-r^2I_n)\right]\nonumber\\
\geq&-\big(T_1^2-T_2^2\bar k_1\big)^2 +4\bar k_1^2\bar k_2^2+16\bar k_1^3-16\bar k_1^2 T_1T_2\nonumber\\
=&4\bar k_1^2\bar k_2^2-T_1^4-T_2^4\bar k_1^2+2T_1^2T_2^2\bar k_1+16\bar k_1^3-16\bar k_1^2 T_1T_2\nonumber\\
>&4T_1^2T_2^2\bar k_1+16\bar k_1^3-16\bar k_1^2 T_1T_2
=  4\bar k_1\left(T_1T_2-2\bar k_1\right)^2\nonumber\\
\geq& 0,
\end{align}
which implies that $V(X)$ is positive definite.
As a consequence, the PD control system  (\ref{sys}) and (\ref{14}) will satisfy (3)
exponentially, for all initial values $x_{1}(0),$ $x_{2}(0)\in\mathbb{R}^{n}$.

\emph{Necessity:} The necessity of Theorem 2 is similar to the proof of Theorem 1(ii), we omit it here due to page limitation.  \hfill$\square$

\section{Conclusion}
This article investigates the capability and limitations of the classical PD control for a class of nonaffine uncertain stochastic systems. We have shown that the nonaffine uncertain stochastic system can be globally stabilized by the PD control in the mean square sense, if the upper bounds of the partial derivatives of the system nonlinear functions satisfy a certain algebraic inequality. Moreover, we have shown that the PD control has fundamental limitations in stabilizing the considered stochastic systems, once the size of the system uncertainty exceeds a critical value. Furthermore, based on some prior knowledge of both the drift and diffusion terms, necessary and sufficient conditions on the selection of the controller gains are also provided for a class of linear uncertain stochastic systems. For further investigation, it would be meaningful to optimize the PD parameters to get better transient performance, and to consider more practical situations including time-delay and saturation, etc.

\section*{Appendix}
\subsection{Auxiliary results}

We provide two lemmas that are used in the proof of the main results.

\begin{Lemma} A necessary and sufficient condition for the set $\Omega$ defined by (\ref{pD}) to be non-empty is
\begin{align}\label{M}4L_1M^4+4N_1M^3+2L_2M^2+2N_2M-1<0.\end{align}
\end{Lemma}
\vskip 0.25cm
\emph{Sufficiency:} First, suppose that (\ref{M}) holds, we will show  $\Omega\neq \emptyset$ by verifying $(k_p^*,k_d^*)\in\Omega$, where
\begin{align}\label{k*}
k_p^*:=\frac{1\!-\!2N_2M\!-\!2L_2M^2\!-\!2N_1M^3}{2M^4},~
k_d^*:=\frac{1\!-\!N_2M}{M^2}.\end{align}
First, note that
\begin{align}\label{k1}
&\bar k_1=k_p^*-L_1\nonumber\\
=&(1-\!2N_2M-2L_2M^2-\!2N_1M^3-\!2L_1M^4)/(2M^4),\end{align}
then it follows from (\ref{M}) that $\bar k_1>0.$
Moreover, it can be obtained that
\begin{align}
\bar k_2=&k_d^*-L_2=\left(1-N_2M-L_2M^2\right)/M^2,\label{422}\\
T_1=&N_1+Mk_p^*=\left(1-2N_2M-2L_2M^2\right)/(2M^3),\label{t1}\\
T_2=&N_2+Mk_d^*=1/M.\label{44}
\end{align}
It follows from (\ref{422})$-$(\ref{44}) that
\begin{align}\label{49}
2\bar k_2-T_2^2=(1-2N_2M-2L_2M^2)/M^2=2MT_1.
\end{align}
From (\ref{k1}) and $(\ref{t1})$,  we obtain the following identity:
\begin{align}\label{59}
&2M^3\left(2M\bar k_1-T_1\right)=4M^4\bar k_1-2M^3T_1\nonumber\\
=&2-4N_2M-4L_2M^2-4N_1M^3-4L_1M^4\nonumber\\
&-(1-2N_2M-2L_2M^2)\nonumber\\
=&1-2N_2M-2L_2M^2-4N_1M^3-4L_1M^4.\end{align}
Thus,  we have $2M\bar k_1-T_1>0$.
Consequently, it follows from (\ref{49}) and (\ref{59}) that
\begin{align}\label{60}
&2\bar k_1\bar k_2-T_1^2-\bar k_1T_2^2=
\bar k_1(2\bar k_2-T_2^2)-T_1^2\nonumber\\
=&2MT_1\bar k_1-T_1^2=(2M\bar k_1-T_1)T_1>0.
\end{align}
From (\ref{k1}) and (\ref{60}), we know that $(k_p^*,k_d^*)\in\Omega$, which implies the non-empty property of $\Omega$.

\emph{Necessity:} Suppose that $\Omega$ is non-empty,
we proceed to show that (\ref{M}) holds.
It suffices to consider the case $M>0$, since (\ref{M}) is automatically satisfied when $M=0$.

Let $\bar \Omega$ be the closure of $\Omega$, i.e.,
\begin{align}\label{empty}
\bar \Omega=\left\{(k_p,k_d)|~ k_p\geq L_1, ~~2\bar k_1\bar k_2-T_1^2-\bar k_1T_2^2\geq 0\right\}.
\end{align}
First, we show that $\bar \Omega$ is bounded(hence it is compact).

Suppose that $(k_p,k_d)\in \bar \Omega$, then
$2\bar k_1\bar k_2-\bar k_1T_2^2\geq 0$, which yields
$2k_d>2\bar k_2\geq T_2^2\geq M^2k_d^2.$ Hence, $k_d<2/M^2$.
Also, from
$2\bar k_1\bar k_2>0$, we know that $\bar k_2>0$, i.e., $k_d>L_2$.

Next, we estimate the bounds of $k_p$. It is easy to obtain
\begin{align}
4k_p/M^2>4\bar k_1/M^2>2\bar k_1 k_d> 2\bar k_1\bar k_2\geq T_1^2>M^2k_p^2,
\end{align}
therefore $L_1<k_p<4/M^4$. Combine this with the bounds of $k_d$, we find that $\bar \Omega$ is bounded.

Define a function $H(\cdot)$ as follows:
$$H(k_p,k_d)=2\bar k_1\bar k_2-T_1^2-\bar k_1T_2^2,~~(k_p,k_d)\in \bar\Omega.$$
By the definition (\ref{empty}) of $\bar \Omega$, we know that $H(k_p,k_d)\geq 0$, for $(k_p,k_d)\in \bar\Omega$ and $H(k_p,k_d)> 0$, for $(k_p,k_d)\in \Omega$.

Since $\bar \Omega$ is compact, we know that $H(\cdot)$ can attain its maximum value.  Note that $H(k_p,k_d)=0$ on the boundary of $\bar \Omega$, thus the maximum point $(k_p^*,k_d^*)\in \Omega$,  and therefore
$\frac{\partial H}{\partial k_p}\big |_{(k_p^*,k_d^*)}=\frac{\partial H}{\partial k_d}\big |_{(k_p^*,k_d^*)}=0$.
By simple manipulations, we have
\begin{align}\label{66}
\frac{\partial H}{\partial k_p}\Big |_{(k_p^*,k_d^*)}=&2\bar k_2-2T_1M-T_2^2=0,\\
\frac{\partial H}{\partial k_d}\Big |_{(k_p^*,k_d^*)}=&2\bar k_1(1-T_2M)=0\label{616}.
\end{align}
%\begin{align*}
%\frac{\partial^2 H}{\partial k_p^2}=-2M^2,~~\frac{\partial^2 H}{\partial k_p\partial k_d}=2-2T_2M,~
%\frac{\partial^2 H}{\partial k_d^2}=-2\bar k_1M^2.
%\end{align*}
It follows from (\ref{66}) and (\ref{616}) that
$$M(N_2+Mk_d^*)=1,~~2(N_1+Mk_p^*)M+(N_2+Mk_d^*)^2=2\bar k_2.$$
Hence, it can be obtained that $k_d^*=(1-MN_2)/M^2$ and $$k_p^*=(1-2MN_2-2M^2L_2-2M^3N_1)/(2M^4),$$
which is exactly the formula given in (\ref{k*}).

Note that $H(k_p^*,k_d^*)>0$, and from (\ref{59})$-$(\ref{60}), we know
\begin{align}
&1-2N_2M-2L_2M^2-4N_1M^3-4L_1M^4\nonumber \\=&2M^3(2M\bar k_1-T_1)=2M^3H(k_p^*,k_d^*)/T_1>0.
\end{align}
Hence, Lemma 1 is proved. \hfill$\square$

Similar to the proof of Lemma 1, we can obtain:
\begin{Lemma} A necessary and sufficient condition for the set $\Omega'$ defined by (\ref{omega'})
to be non-empty is
\begin{align}\label{MM}16L_1M^4+16N_1M^3+4L_2M^2+4N_2M-1<0.\end{align}
\end{Lemma}

\subsection{Proof of Proposition 1.}
Without loss of generality, we assume that $y^*=0$. Suppose that for some $k_p$ and $k_d$ and  for some initial state $(x_1(0),x_2(0))\in\mathbb{R}^n$, the closed-loop equation (1)-(2) satisfies
\begin{align}\label{to0}\lim_{t\to\infty} \mathbb{E} \|x_1(t)\|^2=0 ~~\text{and}~~\lim_{t\to\infty}  \mathbb{E} \|x_2(t)\|^2=0,\end{align}
we proceed to show $f(0,0,0)=g(0,0,0)=0$.

Note that
$u(t)=-k_p x_1(t)-k_d x_2(t)$, it follows from (\ref{to0}) that $\lim_{t\to\infty}\mathbb{E}\|u(t)\|^2 =0.$
Recall	 $\mathrm{d}x_{2}=f(x_1,x_2,u)\mathrm{d}t+g(x_1,x_2,u)\mathrm{d}B_t$,
it follows that
\begin{align}\label{de}
x_2(t+1)-x_2(t)=X_t+Y_t,
\end{align}
where
\begin{align}
&X_t=\int_t^{t+1} f(x_1(s),x_2(s),u(s))\mathrm{d}s,\\
&Y_t=\int_t^{t+1} g(x_1(s),x_2(s),u(s))\mathrm{d}B_s.\end{align}
Next, we proceed to show that
\begin{align}\label{da}
\mathbb{E}\left[X_t^{\mathsf{T}} Y_t\right]\to 0, ~\text{as}~ t\to\infty.
\end{align}
To this end, we first need to prove the following two facts:
\begin{align}\label{466}
&\lim_{t\to\infty}\mathbb{E}\|X_t-f(0,0,0)\|^2=0,\\
\label{46}&\lim_{t\to\infty} \mathbb{E}\big\|Y_t-g(0,0,0)(B_{t+1}-B_t)\big\|^2=0.
\end{align}
%Note that
%\begin{align}\label{126}
%\lim_{t\to\infty}\mathbb{E}[\|x_i(t)\|^2]=0,~i=1,2;~
%\lim_{t\to\infty}\mathbb{E}[\|u(t)\|^2]=0,\end{align}
From the Cauchy-Schwarz inequality, we can obtain
\begin{align}\label{67}
&\lim_{t\to\infty}\mathbb{E}\|X_t-f(0,0,0)\|^2\nonumber\\
=&\lim_{t\to\infty}\mathbb{E}\Big\|\int_t^{t+1} \Big[f(x_1(s),x_2(s),u(s))-f(0,0,0)\Big ]\mathrm{d}s\Big\|^2\nonumber\\
\le&\lim_{t\to\infty}\!\mathbb{E}\int_t^{t+1} \!\|f(x_1(s),x_2(s),u(s))\!-\!f(0,0,0)\|^2\mathrm{d}s.
\end{align}
Moreover, from (\ref{67}) and the Lipschitz property of $f$, we have
\begin{align}\label{45}
&\lim_{t\to\infty}\mathbb{E}\|X_t-f(0,0,0)\|^2\nonumber\\
%\le &\lim_{t\to\infty}\mathbb{E}\int_t^{t+1} \|f(x_1(s),x_2(s),u(s))-f(0,0,0)\|^2\mathrm{d}s\nonumber\\
\le & \lim_{t\to\infty}\mathbb{E}\int_t^{t+1} C\left[\|x_1(s)\|^2+\|x_2(s)\|^2+\|u(s)\|^2\right]\mathrm{d}s\nonumber\\
= & \lim_{t\to\infty}C \int_t^{t+1} \mathbb{E}\left[\|x_1(s)\|^2+\|x_2(s)\|^2+\|u(s)\|^2\right]\mathrm{d}s\nonumber\\
=&0,
\end{align}
for some constant $C>0.$
Hence, (\ref{466}) is proved. Similarly, by the Ito's isometry and the Lipschitz property of $g$, one can prove (\ref{46}) in a similar way.
From (\ref{46}), we know that  $\mathbb{E} \|Y(t)\|^2$ is a bounded function of $t\in[0,\infty)$.

By applying  (\ref{466})-(\ref{46}) again,
it can be obtained that
\begin{align}\label{daaa}
\lim_{t\to\infty}\mathbb{E}\left[X_t^{\mathsf{T}} Y_t-\!f^{\mathsf{T}}(0,0,0)g(0,0,0)(B_{t+1}-B_t)\right]= 0.
\end{align}
On the other hand, note that
\begin{align}\label{daa}
\mathbb{E}\left[f^{\mathsf{T}}(0,0,0)g(0,0,0)(B_{t+1}-B_t)\right]= 0, ~\forall t\geq 0.
\end{align}
Consequently, (\ref{da}) follows from (\ref{daaa}) and (\ref{daa}).

%It comes immediately since for any given $T>0$, we have $$Ef(0,0,u(T))(w(t+1)-w(t))\to 0,~~ t\to \infty$$ and the fact that
% $$\lim_{T\to\infty}f(0,0,u(T))=f(0,0,u_\infty),~~ \text{in}~~ L^2(\Omega,\mathscr{F},P).$$

From (\ref{de}), we know that
\begin{align*}
\mathbb{E}\left[\|X_t\|^2+\|Y_t\|^2+2X_t^{\mathsf{T}}Y_t\right]
=\mathbb{E}\left[\|x_2(t+1)-x_2(t)\|\right]^2.
\end{align*}
Recall $\lim_{t\to\infty}  \mathbb{E} \|x_2(t)\|^2=0$,
we conclude that \begin{align}\label{61}
\lim_{t\to\infty}\mathbb{E}\left[\|X_t\|^2+\|Y_t\|^2+2X_t^{\mathsf{T}}Y_t\right]=0.
\end{align}
From (\ref{da}), we have
 $\lim_{t\to\infty}\mathbb{E}\|X_t\|^2+\|Y_t\|^2=0.$
Combine this with (\ref{466}) and (\ref{46}), we can obtain  $f(0,0,0)=0$ and $g(0,0,0)=0.$ \hfill$\square$

\end{document}